\documentstyle[prl,aps,psfig,floats]{revtex}

\begin{document}

\twocolumn[\hsize\textwidth\columnwidth\hsize\csname@twocolumnfalse\endcsname
\vskip -0.7truecm{\tt To appear on Surface Science}\vskip 0.4truecm

\title{ The mechanism for the $3 \times 3$ distortion of Sn/Ge(111) }

\author{ S.~de~Gironcoli$^{1,2}$, S.~Scandolo$^{1,2}$, G.~Ballabio$^{1,2}$,
         G.~Santoro$^{1,2}$ and E.~Tosatti$^{1,2,3}$ }

\address{
$^{1)}$ International School for Advanced Studies (SISSA)
	via Beirut 4, I-34014 Trieste, Italy\\
$^{2)}$ Istituto Nazionale Fisica della Materia (INFM),
	I-34014 Trieste, Italy \\
$^{3)}$ International Centre for Theoretical Physics (ICTP), 
	I-34014 Trieste, Italy }
\date{\today}
\maketitle

\begin{abstract}
We show that two distinct $3 \times 3$ ground states, one nonmagnetic,
metallic, and distorted, the other magnetic, semimetallic (or insulating)
and undistorted, compete in $\alpha$-phase adsorbates on semiconductor
(111) surfaces.  In Sn/Ge(111), LSDA/GGA calculations indicate, in
agreement with experiment, that the distorted metallic ground state
prevails. The reason for stability of this state is analysed, and is
traced to a sort of bond density wave, specifically a modulation of the
antibonding state filling between the adatom and a Ge-Ge bond directly
underneath.
\end{abstract}

\vspace{0.5cm} 
]
\newpage\narrowtext

\section{Introduction}

The study of the so-called $\alpha$-phases (1/3 monolayer adatom coverage)
of tetravalent adsorbates on semiconductor surfaces has recently attracted
considerable interest, due to the complex and diverse phenomenology
displayed by systems that, at a first glance, look very similar, both from
a structural and from an electronic point of view. On one side we have
Pb/Ge(111) \cite{pb-ge111-stm,pb-ge111-pes,pb-ge111-xrd} and Sn/Ge(111)
\cite{sn-ge111-stm,sn-ge111-pes1,sn-ge111-xrd,sn-ge111-pes2,sn-ge111-pes3,melechko99,zhang99,bunk99}, where a transition from
$\sqrt{3}\times\sqrt{3}$ to $3\times 3$ surface periodicity has been
observed below $\sim 200$ K. On the other side, SiC(0001) \cite{sic-exp}
and K/Si(111):B \cite{si-si111} retain a $\sqrt{3}\times\sqrt{3}$
periodicity at all temperatures, but are insulating, in contrast with
simple electron counting rules. All the above systems are characterized,
in the $\sqrt{3}\times\sqrt{3}$ phase, by a narrow and half-filled surface
band arising from the dangling bond orbital of the adatom. Narrow metallic
bands are highly unstable either against electron-electron instabilities
(Mott transition), or against genuine structural distortions aimed
at lowering the electronic density of states at the Fermi energy
\cite{footnote}. SiC(0001) and, possibly, also K/Si(111):B appear to
belong to the former class, due to the large Coulomb repulsion within the
dangling bond orbital of the Si adatom. In line with this expectation,
SiC(0001) has been recently predicted to be a surface magnetic
Mott insulator based on first-principles calculations \cite{sic-th}.
Sn/Ge(111) and Pb/Ge(111) belong instead to the class of surfaces where
a large structural distortion, leading to a $3\times 3$ periodicity,
removes at least some of the original destabilizing metallic character.

Although considerable progress has been made in the last years,
a complete understanding of the physics underlying the appearence of so
different phenomena (Mott transition or atomic distortion) at the surface 
of otherwise very similar systems, is presently missing. 
In particular, we lack a microscopic understanding of the reasons that 
make a system decide for one or the other state. This is particularly 
relevant also in connection with the possibility that these states may be 
observed at the surfaces of other, presently unexplored, $\alpha$-phases
such as Sn/Si(111), Pb/Si(111). Recent work on Sn/Si(111)
\cite{ottaviano99} fails to indicate any sign of transition down to 100 K.

Here we focus on Sn/Ge(111), as a prototype of the systems where
a large atomic distortion takes place. We investigate the system
with first-principles methods, and find that both an undistorted
(i.e. structurally $\sqrt{3}\times\sqrt{3}$) but magnetic state,
and a $3\times 3$ structurally distorted state, lower the energy of
the originally metallic $\sqrt{3}\times\sqrt{3}$ surface. However, the
energy gain is larger for the structurally distorted case, explaining the
observed low temperature $3\times 3$ structure. Moreover, by examining
in detail the atomic and electronic structure of the $3\times 3$
distortion, we are able to highlight the microscopic mechanisms that
drive the transition.

\section{Method}

The Sn/Ge(111) surface has been modelled in a repeated supercell
geometry where three bilayer slabs of Ge atoms are separated by
equivalently thick vacuum regions. Sn adatoms are placed in the T$_4$
position of the upper surfaces while dangling bonds of the lower
surfaces are saturated by hydrogen atoms.  We performed extensive
electronic structure calculations for both the $\sqrt 3 \times \sqrt
3$ and the $3 \times 3$ surfaces, either in the local (spin) density
approximation (LSDA) or including gradient corrections (GC) \cite{bp}
to the energy functional. Norm-conserving pseudopotential\cite{mt}
in the Kleinman-Bylander form \cite{kb}, a plane-wave basis set with
12 Ry energy cutoff, and a $15 \times 15 $ k-points grid to sample
the full surface Brillouin zone (SBZ) of the $\sqrt 3 \times \sqrt 3$
phase were used. For the  $3 \times 3$ surface an equivalent SBZ sampling
was employed.  All systems were structurally relaxed, keeping the
bottom Ge-layer and the saturating hydrogen atoms fixed, until the
Hellmann-Feynman forces on all other atoms were reduces to less than
$\approx 10^{-2} $ eV/a.u.

\section{Results}

\begin{figure*}[t]
\centerline{
\psfig{figure=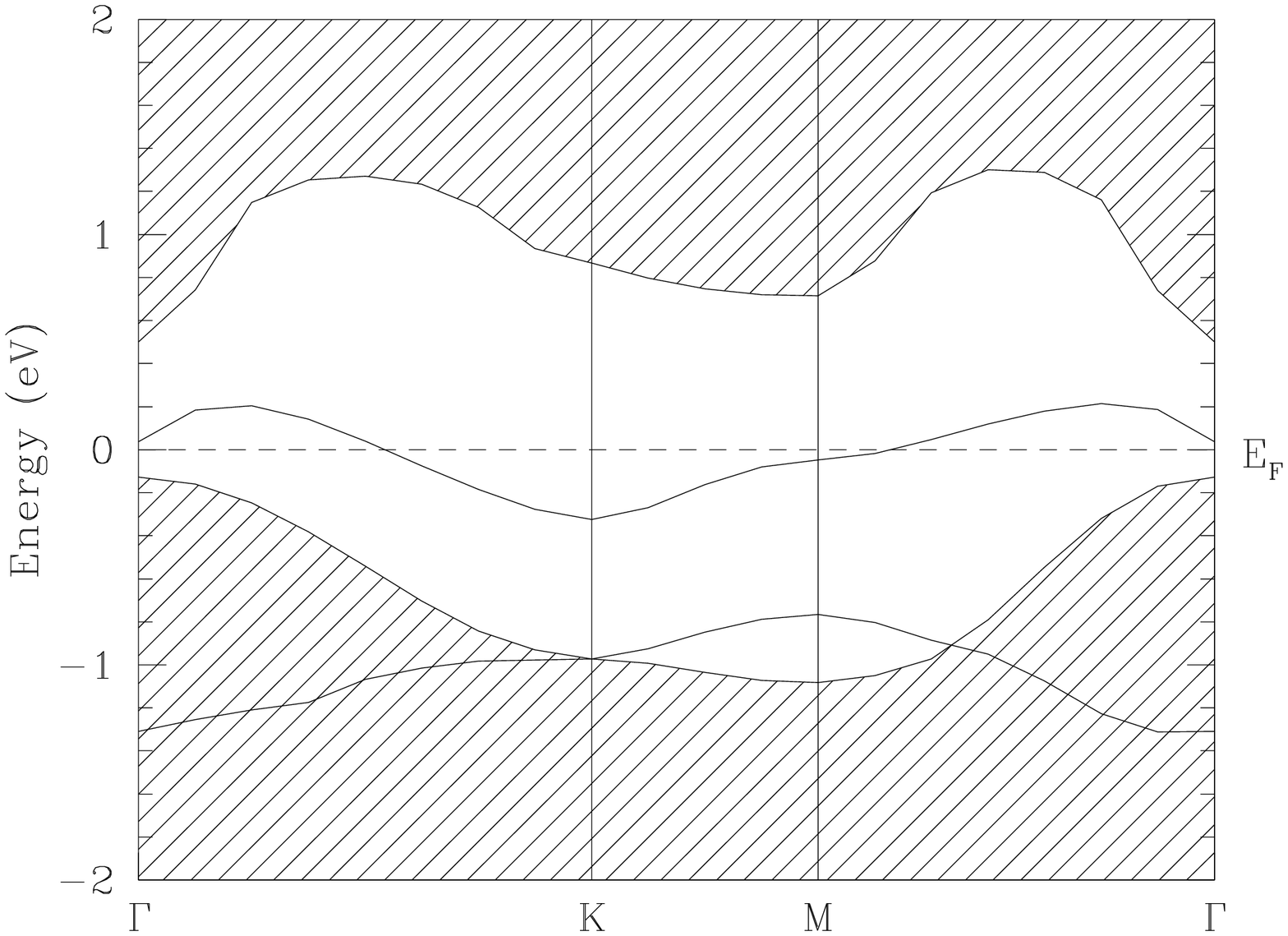,width=7.5truecm}
\psfig{figure=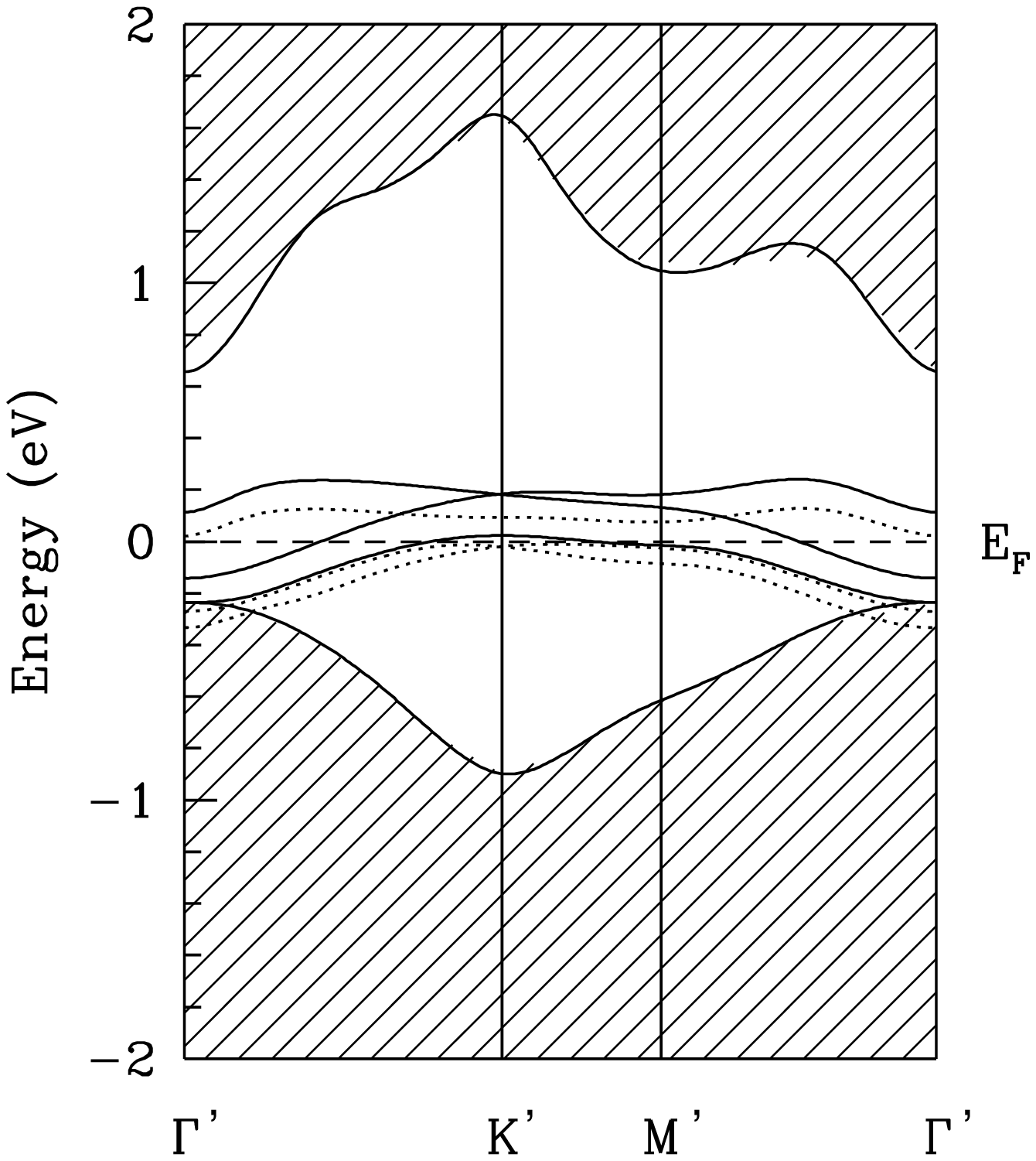,width=5.5truecm}
\psfig{figure=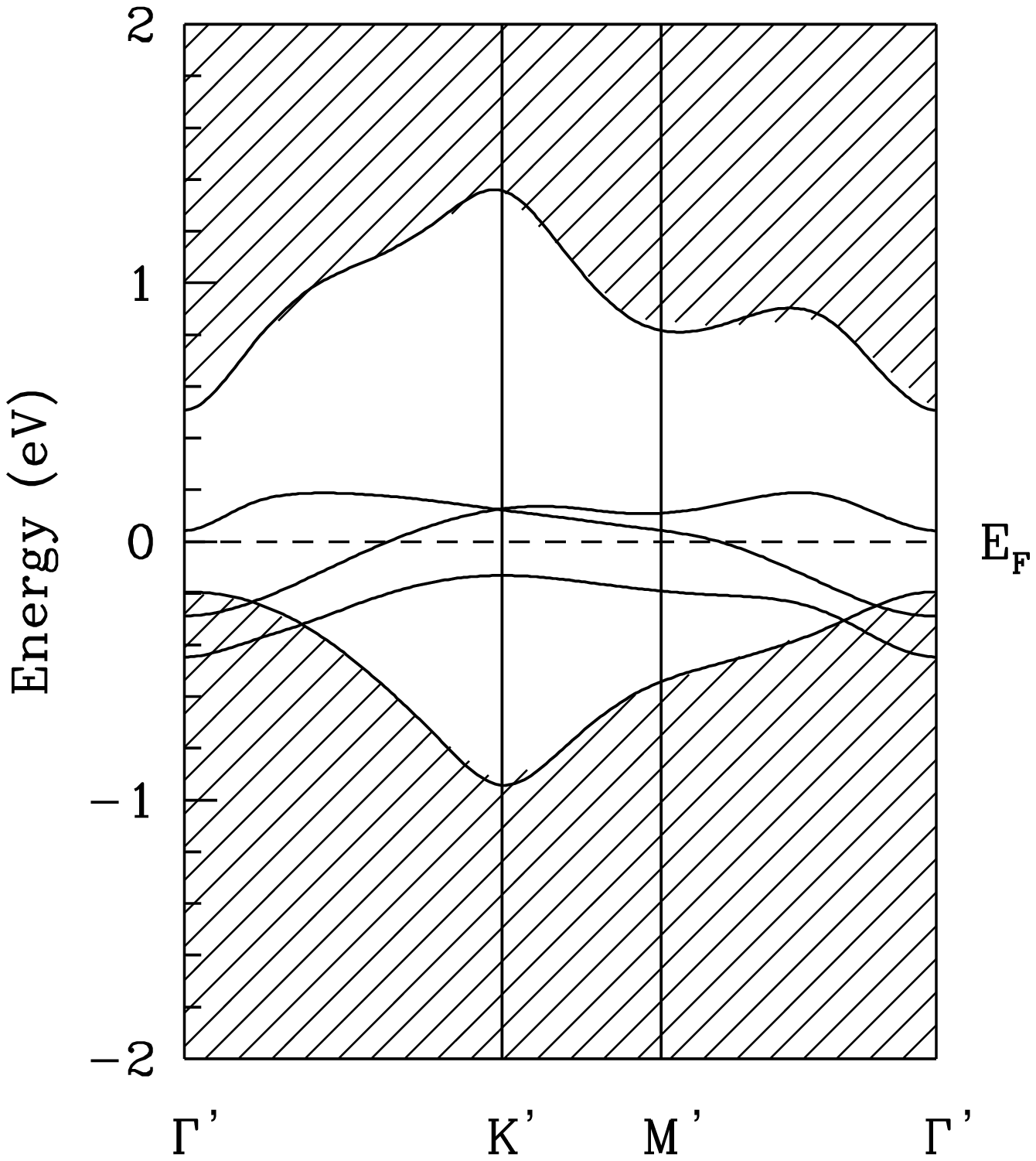,width=5.5truecm}
}
\caption{Surface band structure of the Sn/Ge(111), calculated in the
GC approximation. Left panel: $\sqrt 3 \times \sqrt 3$ undistorted
nonmagnetic surfaces. Middle panel: $3 \times 3$ magnetic surface (here
the structural distortion is negligible). Solid line, spin up electrons;
dotted lines, spin-down electrons. Right panel: $3 \times 3$ distorted
structure (nonmagnetic). In all cases shaded regions represent projected
bulk bands. } 
\label{fig:bands} 
\end{figure*}

The GC band structure of the unreconstructed Sn/Ge (111) $\sqrt 3 \times
\sqrt 3$ surface are reported on the left panel of Fig.\ \ref{fig:bands}.
The system is metallic with a single predominat, partially occupied,
surface band, originating from adatom dangling bonds, in the bulk
projected energy gap.  Due to the rather small bandwidth, $w \approx 0.6$
eV, the system presents an instability toward a spin- and charge-density
wave of $3 \times 3$ periodicity. The corresponding ground-state band
structure is shown in the middle panel of Fig.\ \ref{fig:bands}. In
agreement with experimental evidence, the magnetically ordered structure
is still metallic, although with a reduced density of states at the
Fermi energy. A similar result was obtained earlier for the Si/Si(111)
surface \cite{sacsst-ecoss17} where, however, the exchange gap was
larger and the resulting structure was insulating.  The inclusion in
the calculation of GC terms to the energy functional turns out to be
necessary to stabilize the magnetic structure. This is not surprising
as we expect the underling physics to be rooted in the same aspects
(small band width, large coulomb self-interaction) that produce the Mott
insulating states in related systems\cite{sic-exp,sic-th}.

\begin{figure*}[t]
\centerline{
\psfig{figure=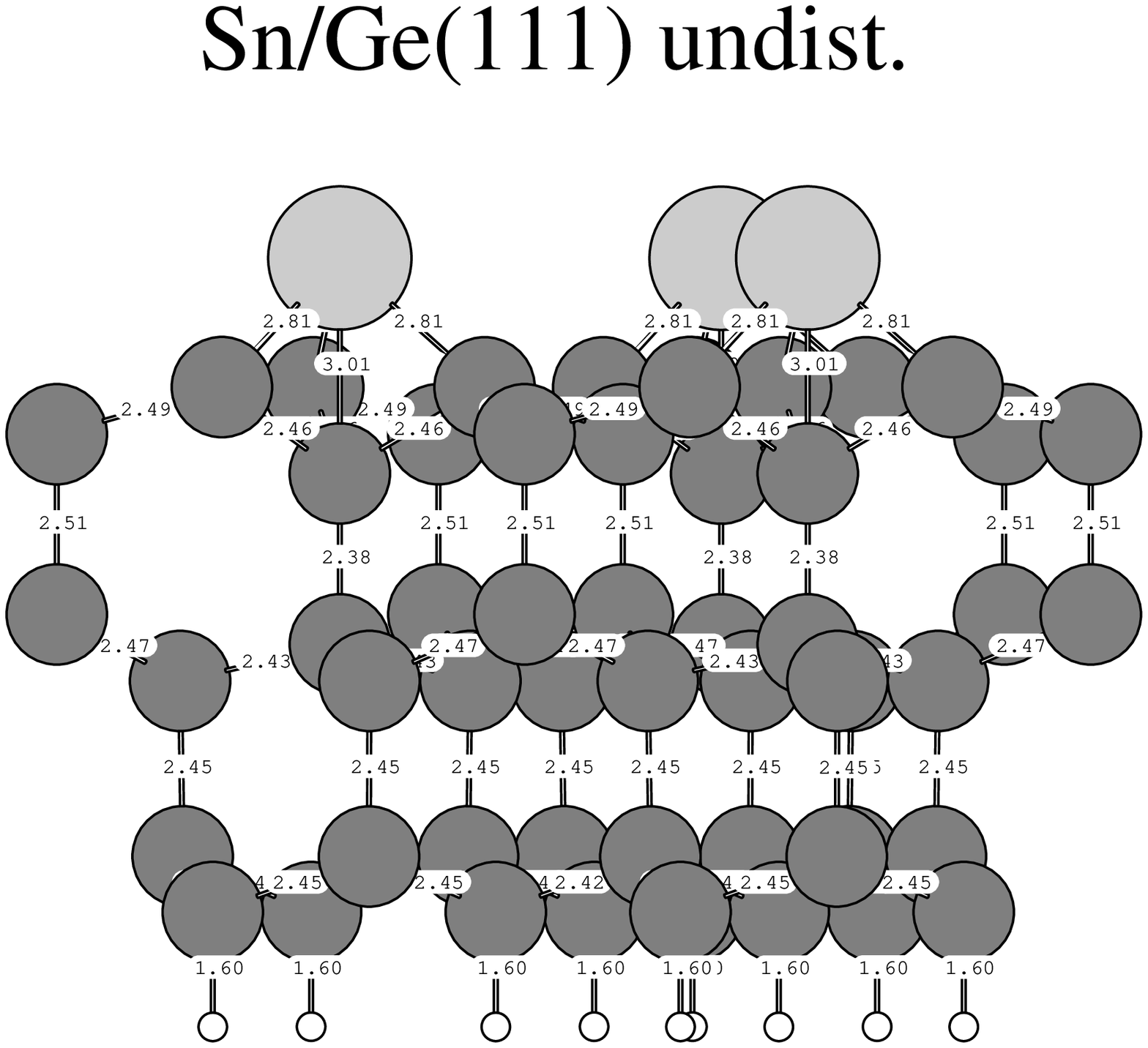,width=8.5truecm}
\psfig{figure=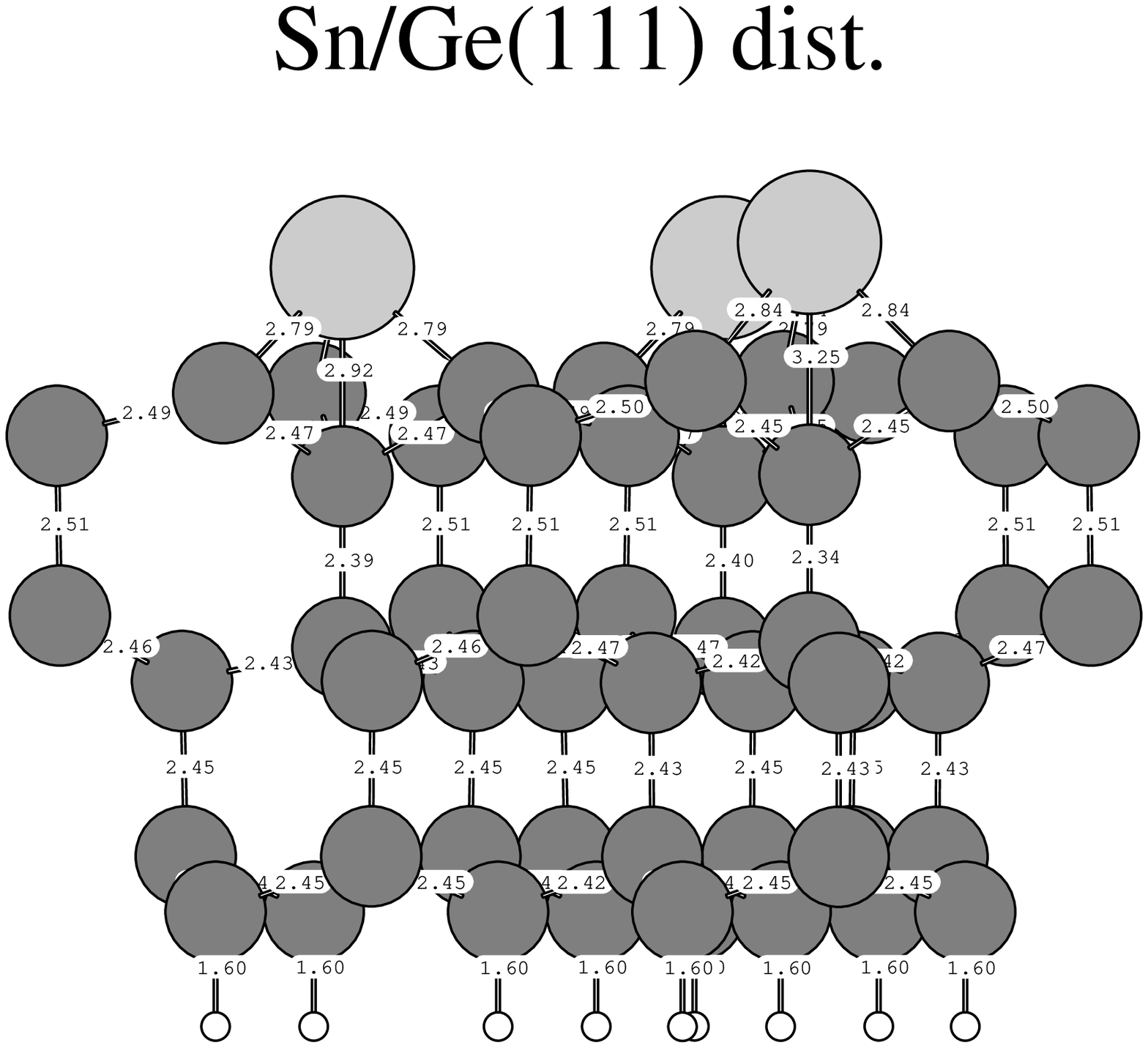,width=8.5truecm}
}
\caption{Atomic geometrical structure of the Sn/Ge(111) surface for
the undistorted ($\sqrt 3 \times \sqrt 3$, left panel) and the distorted 
($3 \times 3$, right panel) structrure. Atomic distances are in \AA}
\label{fig:strut} 
\end{figure*}

In addition to this magnetic instability metallic Sn/Ge(111) $\sqrt 3
\times \sqrt 3$ surface is also unstable against a purely structural
distortion, where vertical displacements of the adatoms with $3 \times 3$
periodicity, are accompained by a bond alternation of the underlying
substrate. In Fig.\ \ref{fig:strut} the atomic structures of the
unreconstructed (left) and reconstructed (right) surface are compared.
One of the 3 adatoms rises above the surface while the other two sink
deeper in the substrate, the final calculated vertical displacement
between the two adatom types being as large as 0.36 \AA. The energy gain
of the system upon reconstruction is of 9 meV/adatom, to be compared with
1-2 meV/adatom of the magnetic case.  The reconstruction is robust with
respect to the exchange-correlation scheme used, and LDA and GC give here
essentially the same results, indicating that the physics involved is
probably well described by a conventional band picture, in spite of
the large value of U/W in the surface state band. This point will be
addressed later. Our calculated distortion compares favorably with that
extracted from very recent x-ray diffraction \cite{bunk99} and also with
a previous LDA calculation \cite{sn-ge111-pes3}.

The reconstruction develops with no energy barrier: we checked that the
energy decreases steadily when the adatom vertical offset is fixed at
about 10 \% of its final value, the substrate atoms being allowed to relax
from the unreconstructed positions.  The role of the substrate relaxation
is important: if only the adatoms positions are optimized, then the energy
gain disappears, and the unreconstructed surface is more stable. Infact,
it can be seen from Fig.\ \ref{fig:strut} that the reconstruction pattern
involves changes in the (111) bonds-lengths between second and third Ge
atomic layers.  The reconstruction apparently occurs because the symmetry
lowering allows some rehybridization between adatoms, accompained by a
bond alternation in the substrate that stabilizes the deformation.

\begin{figure}[ht]
\caption{ 
Contour plot of the mid-gap state at M point in the SBZ of $\sqrt 3 \times
\sqrt 3 $ surface. Full lines enclose regions where the wavefunction is
positive; dotted lines enclose negative regions. Note the antibonding
character of the wavefunction between the adatom and the substrate.
Atomic positions and bonds are also shown for reference.
}
\label{fig:WF} \end{figure}

The mechanism can be better understood by examining the nature of the
Wannier function (WF) of the mid-gap state of the $\sqrt 3 \times \sqrt 3$
surface (approximated here by the wavefunction at the M point, see Fig.\
\ref{fig:WF}). The WF originates from the adatom dangling bond but has
important contributions from substrate states. In particular it can be
seen as an antibonding combination of the adatom dangling bond and the
bonding state located in the (111) Ge-Ge bond below it.  When the system
reconstructs one of the three adatoms becomes inequivalent filling its
state completely. By doing so the corresponding (111) Ge-Ge bond is
strenghtened (due to the bonding character of the WF in that region)
while the Sn-Ge one is  weakened (the WF is antibonding there); as a
result the adatom rises above the surface.  The WF's centered on the other
two adatoms are partially depopulated and an opposite relaxation occurs.

\begin{figure*}[t]
\centerline{
\psfig{figure=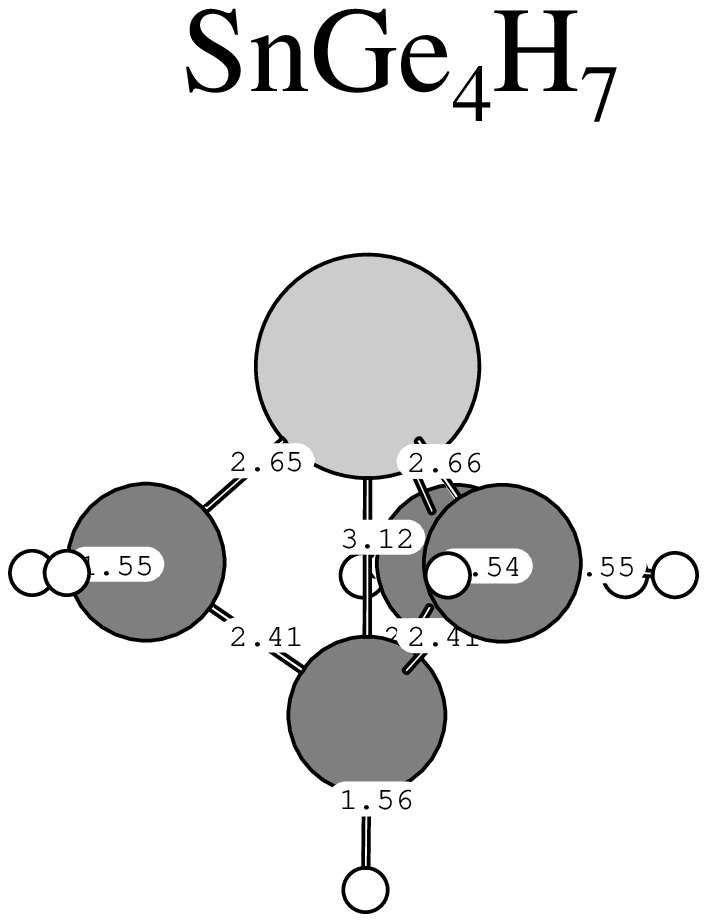 ,width=5.5truecm}
\psfig{figure=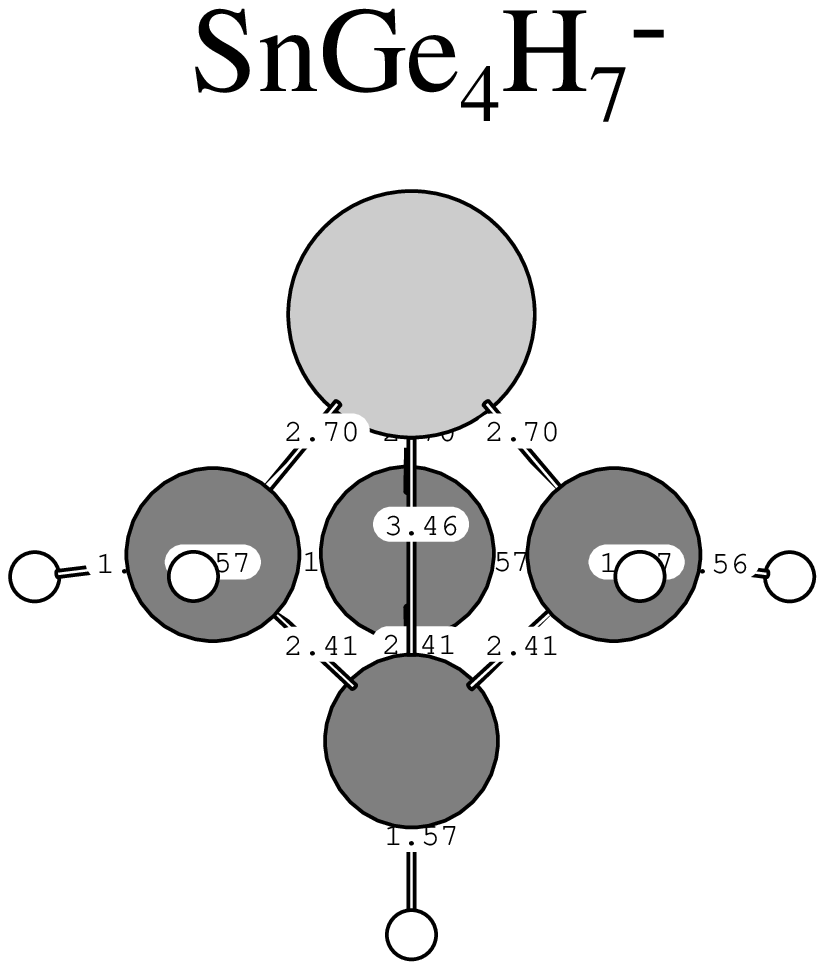,width=5.5truecm}
\psfig{figure=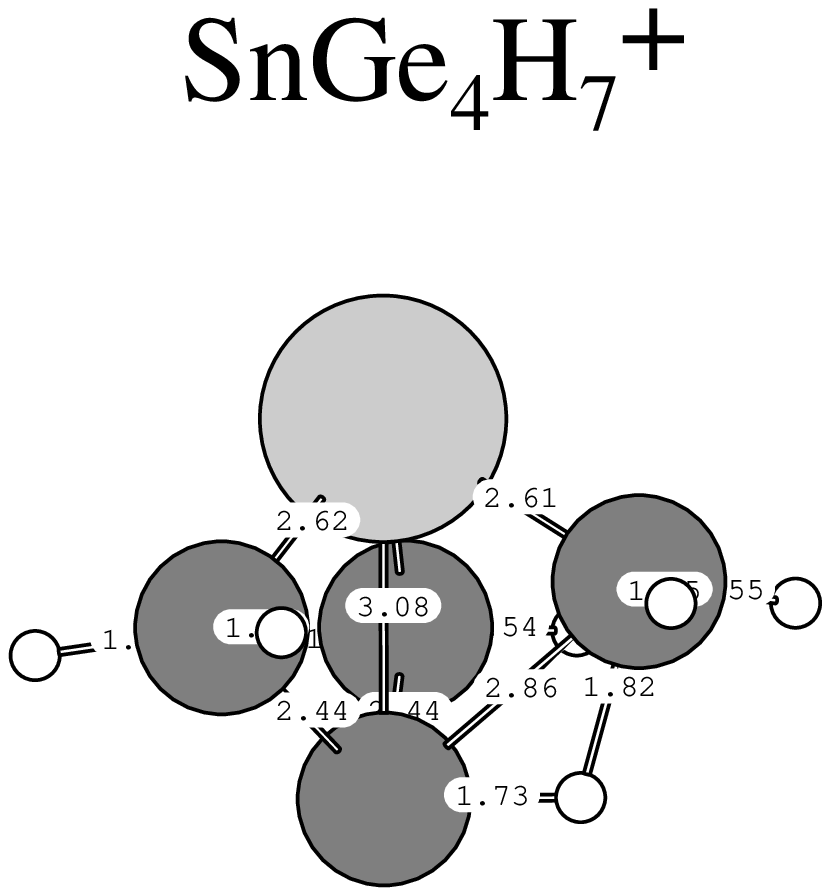,width=5.5truecm}
}
\caption{Ground state geometric structure of a SnGe$_4$H$_7$ cluster
in the neutral (left panel), negatively charged (middle panel) and
positively charged (right panel) states. Atomic distances in \AA. }
\label{fig:mol}
\end{figure*}

To better illustrate this mechanism let us consider a toy system where the
relevant structural motif (a Sn adatom and the four Ge atoms underneath)
is extracted from the surface and hydrogen atoms are added to saturate Ge
dangling bonds.  The neutral cluster (Fig.\ \ref{fig:mol}, left panel)
corresponds to the unreconstructed surface (Fig.\ \ref{fig:strut},
left panel) and presents a semioccupied highest molecular orbital (HMO)
with antibonding character between Sn and Ge.

When the HMO is completely filled, making the cluster negatively charged,
all Sn-Ge bonds weaken and the corresponding distances increase. Not
much else happens; in particular the already strong vertical Ge-H bond
is not modified significantly. The effect is more dramatic in the opposite
situation, when the HMO is emptied making the cluster positively charged
(Fig.\ \ref{fig:mol}, right panel).  As expected from its antibonding
character, depopulating the HMO strenghtens Sn-Ge bonds shortening their
distances, while the lower Ge-H bond is essentially destroyed.

Why does Sn/Ge(111) energetically prefer to distort rather than become
Mott-Hubbard insulating ? Our results suggest that the $3 \times 3$
distortion is based on the strong antibonding interaction between
the adatom dangling bond and the Ge-Ge bond directly underneath.
The fact that the energy gain should come by an alternating
hybridization/dehybridization of the surface band with a deeper Ge-Ge
bond has two implications.  The first is that the large value of U/W {\it
inside} the surface band is not very relevant to this state, unlike the
Mott-Hubbard state. The second is that it shows that band-Jahn-Teller, a
terminology often used to describe it, is not a correct characterization
for this state. It is rather a bond-density-wave. In this case the
energy gain does not come from gap-opening, that is only partial, but
from the modulation of the strength of the Ge-Ge bonds under the adatoms.
Large adatoms, narrow semiconductor gaps and a deformable lattice favor
that. These conditions are not met e.g.\ in SiC(111), where moreover
poor screening enhances the value of electron repulsion (U).

In conclusion, we have studied the competition between magnetic and
distorted nonmagnetic ground states of Sn/Ge(111). Dominance of the
latter has been understood as due to a modulation of the antibonding
partnership between the adatom and the underlying Ge-Ge bond.

Our calculations were performed on the CINECA Cray-T3E parallel machine in
Bologna, using the parallel version of the {\it PWSCF} code.  Access to
the Cray machine has been granted within the {\it Iniziativa Trasversale
Calcolo Parallelo} of INFM and the initiative {\it Progetti di ricerca
di rilevante interesse nazionale} of MURST. Sponsorship from INFM/LOTUS
and from COFIN97 is also acknowledged.


\begin{references}
\bibitem{footnote} In the absence of spin-orbit coupling surface
adatoms in the (3 $\times$ 3) non-collinear Mott magnetic state are
only spin-inequivalent, but charge-equivalent \cite{footref}. Hence,
no periodic lattice distortion is expected in that state.
\end{references}
\end{document}